\documentstyle[sprocl]{article}
\input{psfig.sty}
\def\Journal#1#2#3#4{{#1}{\bf #2}, #3 (#4)}

\def\NPB{{\em Nucl. Phys.} B}
\def\PLB{{\em Phys. Lett.}  B}
\def\PRL{\em Phys. Rev. Lett.}
\def\PRD{{\em Phys. Rev.} D}
\def\ZPC{{\em Z. Phys.} C}

\def\be{\begin{equation}}
\def\ee{\end{equation}}
\def\bea{\begin{eqnarray}}
\def\eea{\end{eqnarray}}

\def \lta {\mathrel{\vcenter
     {\hbox{$<$}\nointerlineskip\hbox{$\sim$}}}}

\begin{document}
\begin{flushright}
MADPH-97-1008\\
~~~\\
~~~\\
\end{flushright}
\title{CONSTRAINTS ON SEMILEPTONIC FORM-FACTORS FROM DISPERSIVE 
SUM RULES
\footnote{Talk presented at the 2nd International Conference on B Physics
and CP Violation, Honolulu, Hawaii, March 24-27, 1997.}
}
\author{G. BURDMAN}
\address{Department of Physics, University of Wisconsin, \\
Madison, WI 53706}
\maketitle\abstracts{
I briefly discuss the derivation of dispersive sum rules
constraining semileptonic form-factors. I outline the use of these
constraints in the context of charmless semileptonic decays and 
suggest how, in combination with other theoretical and experimental 
information, they may reduce the uncertainties in the extraction
of $V_{ub}$.
}
\section{Introduction}
The recent experimental success in observing exclusive charmless  
semileptonic
decays \cite{exp} forces us to reexamine the theoretical situation
in the predictions for these decay modes. 
These seem to
remain plagued with theoretical uncertainties. 
The situation is very different from the one in transitions between hadrons
containing heavy quarks in both the initial and final states, where the 
use of heavy quark effective theory (HQET) leads to very precise 
predictions e.g. for $B\to D^*\ell\nu$ \cite{luke}.
For heavy-to-light transitions
like $B\to\pi\ell\nu$, $B\to\rho\ell\nu$, etc., the impact of the theoretical
developments in heavy quark symmetry is less significant, and is 
mostly reduced to flavor symmetry relations among $B$ and $D$ decay
semileptonic form-factors. There are a variety
of approaches to this problem. In general, complete calculations of the 
form-factors are highly model-dependent. Such is the case with relativistic
and non-relativistic quark models and the various QCD sum-rule calculations.
On the other hand, in some cases model-independent statements can be made 
about these decays. Here we derive one of these model-independent results
from the asymptotic behavior as dictated by perturbative QCD together
with the analyticity of the form-factors \cite{bk}.
As we will see, 
they can have a great impact in the determination of the $q^2$-dependence
of the semileptonic form-factors, a crucial feature in predicting the rate
for these modes and ultimately in extracting $V_{ub}$. 

\noindent
We concentrate on the $B\to\pi\ell\nu$ decay for concreteness and also 
simplicity. 
The hadronic matrix element for the $B^0\to \pi^-\ell^+\nu$ transition can be 
written as 
\be
\langle \pi({\bf p_\pi}) |\bar{u} \gamma_\mu b | B({\bf P})
\rangle = f_+(q^2) (P+p_\pi)_\mu + f_-(q^2) (P-p_\pi)_\mu \label{ff_def}
\ee
where $q^2=(P-p_\pi)^2$ is the momentum transferred to the leptons. 
In the approximation where the leptons are massless, only 
the  form-factor $f_+(q^2)$ enters
the in partial rate. This form-factor obeys a dispersion relation 
of the form
\be
f_+(t)=\frac{-\gamma}{(m_{B^*}^{2}-t)}+\frac{1}{\pi}\int_{s_0}^{\infty}
\frac{Im\left[f_+(s)\right]ds}{(s-t-i\epsilon)} \label{un_dr}
\ee
where $s_0=(m_B+m_\pi)^2$. The isolated pole at $m_{H^*}<s_0$ is actually 
present in the $H=B$ case, whereas for the $D\to \pi$ transition
the $D^*$ pole is located almost exactly at threshold. 
Although not strictly necessary for our results to be derived, it is 
instructive to characterize the different contributions to the imaginary
part in (\ref{un_dr}). These are all possible intermediate states coupling
to $B\pi$ and annihilated by the weak vertex, including the multi-particle
continuum as well as resonances. These must be radial excitations of the 
$B^*$ in order to couple to this current and therefore are expected to be 
separated from the ground state by a typical hadronic scale of approximately
$\simeq 1$~GeV. Thus we expect the continuum to dominate the region just
above the threshold $s_0$, while the resonances take over at a typical
hadronic scale above $s_0$. This picture can be expressed by 
\be
f_+(t)=\frac{-\gamma}{(m_{B^*}^{2}-t)}+
\frac{1}{\pi}\int_{s_0}^{\Lambda^2}\frac{Im\left[f^{\rm cont.}_+(s)\right]ds}
{(s-t-i\epsilon)} + \sum_{i}a_i {\cal R}_i(t) ,  
\label{sep_dr}
\ee
where the cutoff in the integral corresponds to the approximate value of
$s$ where resonances start dominating the imaginary part.
The function ${\cal R}_i(t)$ defined in (\ref{sep_dr}) is generally
given by
\be
{\cal R}_i(t)=\frac{1}{\pi}\left(\frac{\pi}{2}-\arctan{\frac{s_0-M^{2}_i}{
M_i\Gamma_i}}\right) \frac{(M_{i}^{2}-t)}{(M_{i}^{2}-t)^2+M_{i}^{2}
\Gamma_{i}^{2}}~~, 
\label{fw_r}
\ee
whereas it takes the simpler form ${\cal R}_i(t)=(M_i^2-t)^{-1}$ in the 
narrow width approximation. 
It has been common practice in the literature to approximate $f_+(t)$
by the first term in (\ref{un_dr}) and (\ref{sep_dr}). This 
approximation, nearest singularity dominance, is justified in some  cases.
For instance, for values of $t$ very close to  $m_{B^*}$ the first term will
clearly dominate. In the case of the $B\to\pi$ transition, this 
is formalized in the context of chiral perturbation theory for heavy hadrons
(CPTHH) \cite{cpthh}: the $B^*$-pole is the leading contribution in the 
double (heavy quark plus chiral) expansion, which implies that $f_+(t)$ is 
given by the first term in (\ref{un_dr}), 
\be
f_+(t)\approx\left(\frac{g f_B m_B^2}{f_\pi}\right)\frac{1}{m_{B^*}^2-t}~~,
\label{bstpole}
\ee
where the constant $\gamma$ is now given in terms of the $B$ meson
decay constant $f_B$ and the $(H^*-H-\pi)$ coupling $g$. However, this 
expression for the form-factor is only valid for low values of the pion
recoil momentum, where the chiral expansion is justified. The $B^*$-pole
approximation is not to be trusted for pion momenta above a typical hadronic
scale of ($m_\rho - 1$~GeV). Extrapolating the validity of this approximation
is particularly dangerous in $B\to\pi\ell\nu$, where most of the rate tends
to be at large recoil due to the presence of the $|p_\pi|^3$ factor.

\section{Asymptotic Behavior and Dispersive Sum Rules}

The analytic function $f_+(t)$ not only describes the process 
$B\to\pi\ell\nu$ but also processes like s-channel $\ell\nu\to B\pi$
and the t-channel $\ell B\to\nu\pi$. Moreover, its behavior 
for very large values of $|t|$
can be  reliably estimated in perturbative QCD for exclusive 
processes (pQCD) \cite{pqcd}. In this approach the hadronic 
matrix element is described by the hard scattering transition amplitude
folded into an overlap integral between the two hadron state wave-functions.
To leading order, the hard scattering amplitude is approximated by 
the one-gluon exchange diagrams. The gluon momentum satisfies
\be
Q^2=(1-x)^2m_{B}^{2}+(1-y)^2m_{\pi}^{2}\pm 2(1-x)(1-y)P\cdot p_{\pi}
\label{glu_mom}
\ee
where $x$ and $y$ are the momentum fractions of the non-spectator quarks
in the initial and final hadron, respectively. 
The positive sign in the last term in (\ref{glu_mom}) corresponds to
the s-channel process $\ell\nu\to B\pi$, whereas the negative sign 
corresponds to the t-channel, e.g. $\ell B\to \nu\pi$ as well as to the 
$B$ decay. In order for pQCD to be safely applicable we 
need $Q^2\gg 1 {\rm GeV}^2$.  However, the wave-function of a meson 
containing a heavy quark peaks at $x\simeq (1-\epsilon)$, with 
$\epsilon\simeq {\cal O}(\Lambda_{\rm QCD}/m_b)
$.
This implies that in the physical region for the decay $B\to\pi\ell\nu$
the gluon momentum is $Q^2\lta 1{\rm GeV}^2$, with the exception of 
a negligible high-$Q^2$ tail of the wave-function. This casts a serious 
shadow over the applicability of the one-gluon exchange approximation 
in computing $f_+(t)$ {\em in the physical region} for the semileptonic
decay, signaling possible large corrections. 

\noindent
On the other hand, outside the 
physical region and for large enough values of $|t|$, the condition
$Q^2\gg 1{\rm GeV}^2$ is satisfied. 
In these two regions, for $t\ll 0$ or for $t\gg M^2$ 
with $M$ the 
typical mass of heavy resonances, pQCD should yield a 
very good  approximation to the form-factor, especially to its shape. 
This knowledge of $f_+(t)$ 
outside the physical region can be used as a boundary condition
and connected to the decay region by using the dispersion relation 
(\ref{un_dr}). For this purpose, it is convenient to rewrite it 
as
\bea
f_+(t)&=&-\frac{\gamma}{(m_{B^*}^2-t)} + \frac{1}{\pi}\int_{s_0}^{\Lambda^2}
\frac{Im[f_{+}^{\rm cont.}(s)] ds}{s-t-i\epsilon} 
+ \frac{1}{\pi}\int_{\Lambda^2}^{\Lambda'^2}\frac{Im[f_{+}(s)] ds}
{s-t-i\epsilon} \nonumber\\
&+& \frac{1}{\pi}\int_{\Lambda'^2}^{\infty}\frac{Im[f_{+}(s)] ds}
{s-t-i\epsilon}~. \label{dr_asy}
\eea
In (\ref{dr_asy}), the scale $\Lambda'$ defines the end of the 
resonance region
and the beginning of the perturbative regime. For sufficiently large 
$\Lambda'$
the last term can be computed in pQCD. 
For very large values of $|t|$, for instance $t\ll 0$, the asymptotic 
behavior 
of (\ref{dr_asy}) implies
\bea
f_+(t)\longrightarrow \frac{-1}{t}&\times&\left\{-\gamma +\frac{1}{\pi}
\int_{s_0}^{\Lambda^2}
Im[f_{+}^{\rm cont.}(s)]ds \right.\nonumber \\
& &\left.+\frac{1}{\pi}\int_{\Lambda^2}^{\Lambda'^2}
Im[f_{+}(s)]ds +p_1(t,\Lambda') \right\} \label{fp_asy}
\eea
where the last term in the brackets in (\ref{fp_asy}) is the leading 
asymptotic contribution from the last term in (\ref{dr_asy}).
But in this limit we have 
$f_+(t)\longrightarrow f_+^{\rm pQCD}(t)$.
The  non-perturbative
contributions in  (\ref{fp_asy}) are, {\em individually}, much larger
than $f_+^{\rm pQCD}(t)$, which is 
of order $(\alpha_s/\pi)$. Therefore, because $f_+^{\rm pQCD}(t)$
is a reliable approximation to the form-factor for $t\to -\infty$, there
must be large cancellations among the non-perturbative contributions. 
This leads to a convergence condition or dispersive sum rule of the form:
\be
\gamma -\frac{1}{\pi}\int_{s_0}^{\Lambda^2}Im[f_{+}^{\rm cont.}(s)]ds 
-\frac{1}{\pi}\int_{\Lambda^2}^{\Lambda'^2}Im[f_{+}(s)]ds \simeq 0 
\label{con_sr}
\ee
where the equality corresponds to $f_+^{\rm pQCD}(t)=-p_1(t,\Lambda ')/t$.
\footnote{The dependence on the scale $\Lambda'$ reflects the typical scale
dependence of $pQCD$ calculations.}
With this identification, corrections to (\ref{con_sr}) are
of order $(\alpha_s/\pi)^2$. 
This sum rule translates our knowledge of the asymptotic behavior
of $f_+(t)$ into a constraint on the non-perturbative contributions, 
which in turn dominate $f_+(t)$ in the physical region. 
Similar sum rules can be obtained for the other form-factors involved 
in weak decays. The important lesson is that the known 
asymptotic behavior implies a relation among the various non-perturbative
components that make up the transition form-factor. This type of 
relations, as we will
see below, might result in large deviations from the $B^*$-pole behavior
of $f_+(t)$. 

\section{Consequences for $B\to{\rm (light)~}\ell\nu$ Decays}
In order to have an idea of the impact of the dispersive
sum rules on our understanding of the weak form-factors, let us
write (\ref{con_sr}) in the narrow width approximation for the resonances,
\be
\gamma - c - \sum_{i=1}a_i \simeq 0 ~,
\label{con_nwa}
\ee
where $c$ is the integral above threshold of the continuum contribution
and the constants $a_i$ are analogous to $\gamma$ but for the radially
excited states of the $B^*$.  
\noindent First of all, it is not possible to have a large number of 
resonances contributing significantly to (\ref{con_nwa}). The $a_i$'s
are proportional to the couplings between the excited state and the ground
state $(B,B^*)$ and a pion. But the Adler-Weisberger sum rule \cite{aw}
constraints the sum of the squares of all the couplings of excited states
(radial and orbital) to be essentially $1$. Either we have lots of resonances
weakly coupled or only a few with couplings similar to $g$.
For instance, as an exercise one can construct a simple model with two radial 
excitations \cite{bk}. Here it is enough to use 
masses as computed in some successful potential model \cite{potmod} 
given that the intrinsic uncertainties in the spectra derived from these 
calculations will not be of significance in our results. Assuming typical
values for the constants $f_B$ and $g$, making use of the dispersive sum rule
leaves us with one unknown coupling which we fix by fitting to the 
$D^0\to\pi^-\ell^+\nu$ branching ratio. The result is the solid line of 
Figure~1.
\begin{figure}
\hskip 2.60cm
\psfig{figure=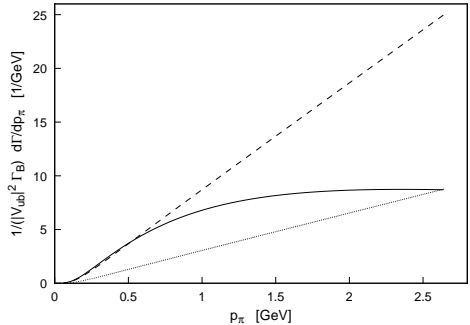,height=1.75in}
\caption{Pion-momentum distributions for $B^0\to\pi^-\ell^+\nu$. 
See text for captions.}
\end{figure}
For comparison, the pure $B^*$-pole prediction is shown in the dashed line
as normalized by CPTHH (\ref{bstpole}). Also, shown by the dotted line, 
is the prediction of the BSW model \cite{bsw}. 
We see that at low pion momentum the model coincides with the $B^*$-pole 
as required
by CPTHH. However, at higher recoils the spectrum softens considerably
due to the imposition of the dispersive sum rule (\ref{con_nwa}), 
which forces the form-factor to behave as dictated by perturbative QCD 
at very large $|t|$. In fact, if one computes $f_+^{\rm pQCD}(t)$ one 
sees that for $t\ll0$, but still 
$|t|<m_B^2 {\rm ln}(m_B^2/\Lambda_{\rm QCD}^2)$, 
the shape of form-factor is given by 
a dipole rather than a monopole\footnote{For 
$|t|\gg m_B^2 {\rm ln}(m_B^2/\Lambda_{\rm QCD}^2)$, the heavy quark mass is 
irrelevant and a monopole behavior similar to that of the pion 
form-factor is recovered.}.
The suppression
with respect to  the $B^*$-pole at large $|p_\pi|$ forced by the sum rule 
gives the correct matching between the two behaviors. 
To see how this happens let us go back to the simple model with a
few narrow resonances. The $B^*$-pole term of the form-factor is now 
modified to be \cite{bk}
\be
\frac{\gamma}{m_{B^*}^2-t}\left(\frac{M_{i}^2-m_{B^*}^2}{M_{i}^2-t}\right)
\simeq \frac{\gamma}{m_{B^*}^2-t}\left(\frac{1}{1+v.p_{\pi}/\Delta_i}\right)
\label{int_sup}
\ee
where $\Delta_i\equiv M_i-m_B$ is the gap between the ground state 
and the {\em i}-th excitation, and provides the scale of suppression 
which is of about $(0.8-1)$~GeV, a typical hadronic scale. 
Therefore the resonances
in the cut give the scale for the effective suppression of the $B^*$-pole 
behavior at large pion recoil. This picture also  provides us with a 
rule of thumb for the validity of the nearest-pole-dominance approximation.
This will be valid as long as the hadronic recoil is small compared to 
the gap between the dominant resonance (e.g. the $B^*$) and the next 
excitation with the right quantum numbers. This is the reason why 
pole dominance is a good approximation in $D$ semileptonic 
decays as well as in $B\to D^{(*)}\ell\nu$ decays, but not 
in $B\to\pi\ell\nu$, where large portions of the rate come from 
$v.p_{\pi}>\Delta_i$.

\section{Outlook}
\noindent
Dispersive sum rules are an important constraint on model building,
as the very simplistic models presented here show.
Although these models are useful as long as they are a correct
parametrization of the physics, more sophistication is possible and may even
be required.
Another step in this approach is to apply it 
to other
charmless decays, e.g. $B\to\rho\ell\nu$, and build similar models
constrained by them that can be tested against experiment.  
The presence of a vector meson in the final state implies that there
are now five independent form-factors, although 
only four are relevant if $m_\ell=0$. 
Recently new relations among heavy-to-light
form-factors have been derived \cite{stech}. In principle these relations
can be used in a region of phase space that might be of interest, and in 
combination with the methods I described earlier yield predictions for 
decay modes like $B\to\rho\ell\nu$ \cite{inprep}. 

\noindent 
Other model-independent avenues, like lattice calculations \cite{jim}
or their combination with the bounds derived from perturbative QCD
dispersion relations \cite{bgl,ll} already place stringent constraints
on the form-factors. 
A complete program incorporating all model-independent
constraints seems to be already emerging and might result in 
theoretically safe predictions for the heavy-to-light form-factors and 
a precise determination of $V_{ub}$.

\section*{Acknowledgments}
The work reported in here is based on a collaboration with J. Kambor
\cite{bk}.
It was supported  by the U.S.~Department of Energy 
under  
Grant No.~DE-FG02-95ER40896 and the University of 
Wisconsin Research Committee with funds granted by the Wisconsin 
Alumni Research Foundation.

\end{document}